\documentstyle[aps,epsf,floats,twocolumn,prb]{revtex}

\def\etal{~\textit{et~al.}} 
\def\ra{\rangle} 
\def\la{\langle} 
\newcommand{\mb}[1]{{\mathbf #1}} 
\newcommand{\bfgr}[1]{{\boldmath{\mbox{$#1$}}}}

\def\undertext#1{\vtop{\hbox{#1}\kern 1pt \hrule}}

\def\ua{\uparrow}
\def\da{\downarrow}

\def\be{\begin{equation}}
\def\ee{\end{equation}}
\def\bea{\begin{eqnarray}}
\def\eea{\end{eqnarray}}

\begin{document}
\twocolumn[\hsize\textwidth\columnwidth\hsize\csname@twocolumnfalse%
\endcsname
\draft

\title{Microscopic models for fractionalized phases in strongly 
correlated systems}
\author{T. Senthil and O. Motrunich}
\address{Massachusetts Institute of Technology,
77 Massachusetts Ave.,  Cambridge, MA 02139
}

\date{\today}
\maketitle

\begin{abstract}
We construct explicit examples of microscopic models that stabilize a 
variety of fractionalized phases of strongly correlated systems in 
spatial dimension bigger than one, and in zero external magnetic field. 
These include models of charge fractionalization in boson-only systems, 
and various kinds of spin-charge separation in electronic systems. 
We determine the excitation spectrum and show the consistency with 
that expected from field theoretic descriptions of fractionalization. 
Our results are further substantiated by direct numerical calculation
of the phase diagram of one of the models.

\end{abstract}
\vskip 0.3 truein
]


\section{Introduction}
Considerable theoretical effort has gone into understanding 
the possibility of obtaining fractional quantum numbers for 
the excitations of strongly correlated systems in two or more 
spatial dimensions, and in weak or zero magnetic fields.
Though much of the original interest arose in 
theories\cite{PWA,KRS,z2long} of the high temperature superconductors, 
ideas based on fractionalization have since been proposed to account for 
the properties of a number of other poorly understood strongly correlated 
systems\cite{coldea,2dwcr,lhull,sudip,marst1,marst2}. 
Field theoretic methods have enabled enormous progress in obtaining 
a description of fractionalization\cite{RSSpN,Wen1,NLII,z2long}.  
A number of exotic  fractionalized phases have been accessed.  
The structure of the distinct possible excitations and the effective 
theory of their interactions\cite{z2long} has been elucidated in some 
detail.  A crucial feature is the presence of discrete gapped vortex-like 
excitations - dubbed visons - apart from the excitations with fractional 
quantum numbers. 
It has become clear\cite{Wen1,topth} that fractionalized phases 
may be given a precise theoretical characterization through the notion of 
``topological order''---a concept elucidated clearly by Wen\cite{Wen2} 
in work on the quantum Hall effect. 

Scepticism has been voiced in some quarters over these developments due to 
the almost complete lack of microscopic models that can be shown to 
display the phenomena mentioned above.  Specifically, consider a model 
of a many-particle system with short-ranged interactions, and no special 
symmetries other than the global charge and/or spin conservation. 
Can fractionalization be shown to obtain in such a model?  
Apart from it's conceptual value, answering this question will also help 
clarify the nature of the microscopic conditions that make it favourable 
for fractionalization to occur in a strongly correlated system. 

There has been limited (though important) progress in a direct answer to 
this question.  Numerical studies\cite{lhull} of a particular triangular 
lattice quantum spin-$1/2$ Heisenberg magnet with ring exchange 
interactions provide evidence for a state with a spin gap and a
four-fold degenerate ground state on a torus as expected in a 
topologically ordered ``spin-liquid'' with fractionalized ``spinon'' 
excitations.  In the context of quantum dimer models\cite{qdm}, 
Moessner and Sondhi\cite{MoeSon_qdm} have argued for a 
stable topologically ordered ``liquid'' phase on a triangular lattice.  
The standard interpretation of the dimer model views the dimers as 
caricatures of singlet bonds formed between underlying Heisenberg spins 
on the lattice.  From this point of view, the work of 
Ref.~\onlinecite{MoeSon_qdm} provides supporting evidence, though not 
definitive proof, that models of Heisenberg antiferromagnets on triangular 
lattices do support a fractionalized spin liquid phase. 
However, as is well-known\cite{FrKiv}, the quantum dimer model is exactly 
equivalent to a gauge theory---thus one may worry that establishing a 
topologically ordered phase in the dimer model still does not convince 
a sceptic that such phases can result in microscopic models with no 
special symmetries or a gauge structure. 

Recently, Balents\etal\cite{BalMPAFGir} argued that a particular 
easy axis quantum spin-$1/2$ model on a Kagome lattice with short ranged 
(albeit complicated) interactions has a topologically ordered ground 
state with fractionalized excitations.
This was done by reinterpreting it as a soluble point of the 
quantum dimer model on a triangular lattice but with three rather than 
one dimer emerging from each site, and following the same arguments as in 
Ref.~\onlinecite{MoeSon_qdm}. 
(See also Ref.~\onlinecite{NaySht} for a somewhat similar perspective.)
However, some features of this model such as the presence of two distinct 
visons, appear to be non-generic (from the point of view of the effective 
field theory of fractionalized phases). 
This model also has an infinite number of local symmetries, 
and hence violates the requirement that fractionalization be demonstrated 
in models with no special symmetries. 
However, Balents\etal\cite{BalMPAFGir} made the important observation that 
perturbing the model slightly to get rid of the local symmetries will
preserve the fractionalized phase. 

In this paper, inspired by these prior developments, we explicitly 
construct microscopic models that stabilize a wide variety of 
fractionalized phases.  These include models for charge fractionalization 
in boson-only systems and various kinds of spin-charge separation in 
electronic systems.  Our models involve only short-ranged interactions, 
and do not have any special symmetries other than global charge and/or 
spin conservation.  We determine the excitation spectrum in the 
fractionalized phases and explicitly show the consistency with that 
expected from the effective field theories. 
Our results complete the answer to the question of principle posed above, 
and will hopefully guide efforts to find materials and other even simpler 
models that realize these phases. 

We begin by illustrating our construction with a simple Bose-Hubbard type 
model of a system of bosons (with charge $q_b$) with short ranged 
interactions.  The model has a global $U(1)$ symmetry 
reflecting the conserved total boson number. 
We explicitly demonstrate the presence of two distinct Mott insulating 
phases in this model.  In one, the excitations carry charges that are 
integer multiples of the underlying boson charge $q_b$. 
In the other, there are excitations with boson number $q_b/2$, 
{\em i.e.}~the bosons have fractionalized.  This phase also has discrete 
$Z_2$ vortices, the visons, which are gapped.  
Upon tuning a parameter in the model, it is possible to drive transitions 
from either Mott insulating phase to a superfluid phase. 
We further substantiate our arguments by performing a quantitative 
numerical calculation of the phase diagram of this model. 
The presence of topological order in one of the Mott insulating phases 
is detected numerically by the flux-trapping experiment discussed in 
Ref.~\onlinecite{toexp}.
We explicitly derive the effective field theory of the fractionalized 
phase and show that it is a theory of bosonic charge $q_b/2$ fields 
coupled to a $Z_2$ gauge field in it's deconfined phase.
 
Next, we consider models of electrons coupled to superconducting 
phase fluctuations.  These may be thought of as models of charge $e$ 
electrons interacting with spinless charge $2e$ bosonic Cooper pairs. 
We show how the boson only models above may be extended to include 
coupling to electrons to provide a realization of various spin-charge 
separated phases.  These models therefore provide explicit realizations 
of the routes explored in Ref.~[\onlinecite{NLII,z2long}] for 
spin-charge separation. 
These spin-charge separated phases have spin-$0$ charge $e$ bosonic 
excitations (chargons or holons), spin-$1/2$ charge neutral fermionic 
spinons, and a gapped vison. 

Of special interest are models that stabilize the 
nodal liquid\cite{NLII,z2long}
(alias $d_{x^2 - y^2}$ RVB) phase---this has gapless fermionic 
nodal spinons, and has played an important role in theories of 
the cuprate materials.  While recent experiments\cite{Bonn,Wynn} are not 
very encouraging on the possibility of fractionalization in the cuprates, 
it still is of theoretical interest to demonstrate models that realize 
the nodal liquid phase. 
Another theoretically controversial possibility is that of ordered 
magnetic phases that nevertheless are spin-charge separated. 
This was also first discussed\cite{NLII,topth} in the context of 
cuprate physics but is possibly relevant to a variety of other systems.  
We show how a model that stabilizes such ordered magnetic 
fractionalized phases may readily be obtained. 
This settles any doubts that may have been harbored on the possibility 
of such coexistence between magnetism and fractionalization. 
We then conclude with a brief discussion.

\section{Fractionalization in boson only models}
\label{Sec:boson}
\subsection{Model and general arguments}
\begin{figure}
\epsfxsize=2.0in
\centerline{\epsffile{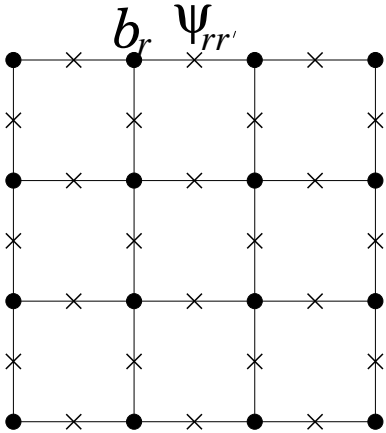}}
\vspace{0.15in}
\caption{Face centered square lattice on which our model 
Eq.~(\ref{H_boson}) is defined.
}
\label{lattice}
\end{figure}

Consider a system of bosons on the ``face-centered'' square
lattice in $2$D shown in Fig.~\ref{lattice} modeled by the Hamiltonian
\begin{eqnarray}
\label{H_boson}
H & = & H_w + H_{\rm bond} + H_{\rm ring} + H_u ~,\\
H_w & = & - w \sum_{r, r'\in r} 
  (b_r^\dagger \Psi_{rr'}+ h.c.) ~, 
\nonumber \\ 
H_{\rm bond} & = & - J_{\rm bond} \sum_{\la rr' \ra} 
  \left[ (\Psi_{rr'}^\dagger)^2 (b_r b_{r'}) + h.c. \right] ~, 
\nonumber \\
H_{\rm ring} & = & - K_{\rm ring} \sum_\Box 
  (\Psi_{12}^\dagger \Psi_{23} \Psi_{34}^\dagger \Psi_{41} + h.c.) ~,
\nonumber \\
H_u & = &  u_b \sum_r (n_r^b)^2
    + u_\psi \sum_{\la rr' \ra} (n_{rr'}^\psi)^2 
    + U \sum_r N_r^2 ~.  \nonumber
\end{eqnarray}

Here $b_r^\dagger = e^{i\theta_r}$ are bosons residing on the corner
sites of the lattice, and $\Psi_{rr'}^\dagger=e^{i\phi_{rr'}}$ 
are bosons on the bond-centered sites, which we identify by 
the end-points of the corresponding bond; $n_r^b$, $n_{rr'}^\psi$
are the corresponding boson numbers, 
$[\theta_r, n_{r'}^b]=i\delta_{rr'}$,
and similarly for $\Psi_{rr'}$ and $n_{rr'}^\psi$. 
For technical convenience, we have chosen a rotor representation of 
the bosons (though this is not essential). 
The operator $N_r$ is defined through 
\begin{equation}
N_r = 2n_r^b + \sum_{r'\in r} n_{rr'}^\psi ~.
\end{equation}
The total boson number of the system is given by 
\begin{equation}
N_{\rm tot} = \frac{1}{2} \sum_r N_r ~.
\end{equation}

The $w$-term is a boson hopping between the corner and the 
bond-centered sites, and $r'\in r$ sums over all such bonds 
emanating from $r$.  The term $K_{\rm ring}$ is a ring exchange among 
four bond-centered sites belonging to the same square plaquette~$\Box$,
while the term $J_{\rm bond}$ is a similar ring-exchange-like boson 
interaction but among three sites associated with a given bond 
$\la rr' \ra$.
The importance of ring exchange terms for promoting fractionalization is 
strongly suggested by the various field theoretic 
descriptions,\cite{z2long}
and by previous studies of microscopic 
models.\cite{lhull,BalMPAFGir}
The $u_b$ and $u_\psi$ are the usual on-site Hubbard terms. 
We have also included the Hubbard-$U$ term for the boson number $N_r$.

\narrowtext

Despite the possibly unfamiliar form of the terms in the Hamiltonian, 
the following features are apparent.  The model clearly has a global 
$U(1)$ charge conservation symmetry associated with global phase 
rotation of all the bosons. 
Note that if the $b$-bosons are assigned charge $q_b$, then the 
$\Psi$-bosons also have charge $q_b$.
There are no other special symmetries for general values of the 
parameters.  In particular, there are no {\em local} symmetries.
Furthermore, all the interactions are {\em short-ranged}. 
We argue below that this model has a stable fractionalized 
insulating phase with charge $\frac{q_b}{2}$ excitations and 
charge $0$ visons above a ground state with no conventional broken 
symmetries. 

Some gross features of the model can be guessed easily. 
At large $w$, the boson kinetic energy dominates and the system will 
be a superfluid.  As $w$ is reduced, there will be a transition to 
an insulating phase.  The nature of this insulating phase depends on 
the other parameters in the model.  In particular, the insulator will
be fractionalized for $J_{\rm bond}, K_{\rm ring}, U$ large compared to 
$u_b, u_\psi$.  In the opposite limit, a conventional 
Mott insulator with charge quantized in units of $q_b$ will obtain. 

\begin{figure}
\epsfxsize=2.7in
\centerline{\epsffile{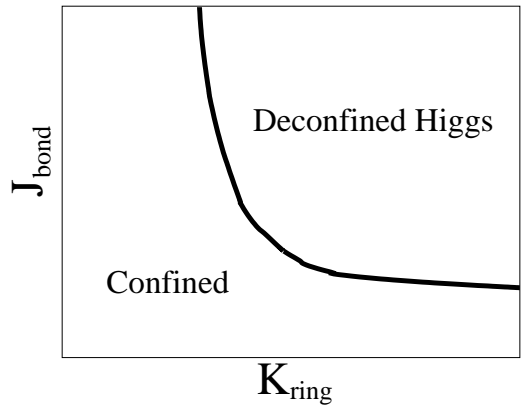}}
\vspace{0.15in}
\caption{Phase diagram of the $(2+1)$D compact QED coupled to a 
charge 2 scalar.
}
\label{QEDc2_phased}
\end{figure}

To establish these results, it is useful to consider the special limit 
$w = 0$; in this case, $[N_r, H]=0$ for every site $r$, 
and we can fix the value of $N_r$ for every $r$. 
Thus, in this limit, the model does have an infinite number of local 
symmetries.  Later, we will move away from this special limit thereby 
destroying these local symmetries.
For large $U$ at $w = 0$, the ground state has $N_r=0$ everywhere.
The $w = 0$ model in the sector $N_r=0$ for every $r$ is readily
understood as it can be regarded as the well-studied\cite{FraShe}
$3$D compact $U(1)$ gauge theory coupled to a charge 2 scalar 
field.  Indeed, divide the underlying square lattice 
of Fig.~\ref{lattice} into $A$ and $B$ sublattices.
Let $\theta_r \to \tilde\theta_r = \epsilon_r \theta_r$ with
\begin{eqnarray}
\epsilon_r & = & +1 \quad {\rm if} \quad r \in A ~,\\
& = & - 1 \quad {\rm if} \quad r \in B ~.
\end{eqnarray}
To preserve the commutation relations, define the corresponding 
conjugate variables
\begin{equation}
\tilde n_r^b = \epsilon_r n_r^b ~.
\end{equation}
Similarly, let $a_{rr'}=\phi_{rr'}$ if $r\in A, r'\in B$
and $a_{rr'}=-\phi_{rr'}$ if $r\in B, r'\in A$.  Consider
$a$ as a vector field $a_{r\alpha} \equiv a_{r, r+\hat\alpha}$,
with $\hat\alpha = \hat x$, $\hat y$, and perform the corresponding
transformation $n_{rr'}^\psi \to E_{r\alpha}$ to the vector field
$E_{r\alpha}$ conjugate to $a_{r\alpha}$.  We have
\begin{equation}
N_r = \epsilon_r\left(\bfgr \Delta \cdot \mb E + 2\tilde n_r^b \right) .
\end{equation}
The Hamiltonian then becomes
\begin{eqnarray}
H & = & - 2 J_{\rm bond} \sum_{r, \alpha} 
\cos(\Delta_\alpha \tilde\theta_r + 2 a_{r\alpha})
- 2 K_{\rm ring}\sum_\Box \cos(\bfgr\Delta \times \mb a) \nonumber \\
& + &  u_b \sum_r (\tilde n_r^b)^2
+ u_\psi \sum_{r, \alpha} (E_{r\alpha})^2,
\end{eqnarray}
while the constraint $N_r=0$ is simply the `Gauss law'
\begin{equation}
\bfgr\Delta \cdot \mb E + 2\tilde n_r^b = 0 ~.
\end{equation}
As promised, $H(w=0)$ is the same Hamiltonian as for 
the $(2+1)$D compact QED coupled to a charge 2 scalar. 
This permits us to take over the classic results of Fradkin and Shenker 
on this model which determined the phase diagram to be of the form 
shown in Fig.~\ref{QEDc2_phased}.
In the ``confined'' phase, all excitations carrying
``gauge charge'' are confined.  In the ``deconfined Higgs'' 
phase, static external objects with gauge charge 1 are not
confined.  Furthermore, there is a stable gapped 
$Z_2$ vortex (which we may identify with the vison). 
A number of different perspectives are available on these results. 
A useful physical one is to regard the deconfined Higgs phase as a 
``condensate'' of the charge-$2$ scalar. 
Naively, such a condensate will have gapped vortices quantized in units of 
$\pi$.  However, due to the compactness of the gauge field, 
space-time monopoles are allowed in the theory. 
These correspond to events where the vorticity changes by 
$2\pi$ - consequently the vortices acquire a $Z_2$ character.   

It is also clear that the deconfined Higgs phase has a topological order:
e.g., the ground state is four-fold degenerate on a torus. 
These are simply obtained by threading no or one vison through the 
two holes of the torus. 

\narrowtext

Consider now the excited states of the original Hamiltonian $H(w=0)$
for large but finite $U$.  Consider states such that $N_{r_0} = 1$ 
at some site $r_{0}$ and $N_r=0$ everywhere else.  Such a state
can be regarded as a static gauge charge $+1$ at $r_0$ 
(assuming $r_0 \in A$).  In the confined phase this
sector costs an infinite energy in an infinite system.
However, in the deconfined Higgs phase it costs only a finite 
energy.  Remarkably, in the original boson model, such a state has a 
true electric charge of 
$\frac{q_b}{2}$ (recall that $Q_{\rm tot} = \frac{q_b}{2} \sum_r N_r$).
Thus, in the deconfined Higgs phase, excitations with fractional 
quantum numbers for the true electric charge are allowed.  
In contrast, in the confined phase, finite energy excitations
have gauge charge $0$ --- this requires that 
$\sum_{r \in A} N_r = \sum_{r \in B} N_r$. 
Consequently, the excitations carry true electric charge that are 
integer multiples of $q_b$, and hence are not fractionalized.

These results on the $w = 0$ Hamiltonian thus follow as a 
straightforward application of the standard Fradkin-Shenker analysis of 
the phase diagram of gauge theories.  
However, they acquire even further importance here when we consider the 
Hamiltonian away from the $w = 0$ limit when $H$ no longer has an 
infinite number of local symmetries. 
Consider small $w$.  This introduces fluctuations which mix states 
with different values of $N_r$ at the same site.  However, for small 
$w$, these will not be capable of closing the gap to excitations about 
the ground state.  Consider in particular the deconfined Higgs phase in 
the presence of a small $w$.  The fractionally charged excitations are 
now allowed to hop from site to site and will acquire a kinetic energy of 
order $w$.  However they will survive as meaningful excitations. 
The other independent excitation, namely the $Z_2$ vortex, will also 
survive introduction of a small $w$.  Thus, the original model has 
for $w \neq 0$ but small a genuine fractionalized phase.  
(We can also add other more general boson hopping terms; clearly,
the fractionalized phase will survive as long as these terms are weak.)
In the subsequent subsections, we provide several direct confirmations 
of these arguments.  In particular, we provide an explicit derivation of 
the effective field theory of the fractionalized phase and show that it is 
a theory of charge $\frac{q_b}{2}$ chargons coupled to a $Z_2$ gauge 
field in it's deconfined phase.  This will also serve to make obvious
our assertions on the properties of the model. 

We emphasize that despite the ease with which this result has been 
obtained, it has enormous significance. 
The Hamiltonian for $w \neq 0$ has no special symmetries other than 
global charge conservation, and has only short ranged interactions. 
Nevertheless, it possesses a fractionalized phase with charge 
$\frac{q_b}{2}$
excitations and a gapped vison consistent with that expected from earlier 
field theoretic descriptions of fractionalization.

\subsection{Numerical calculation of phase diagram}
In this subsection, we substantiate our results by a direct numerical 
calculation of the phase diagram of the model.  To that end, 
it is useful to first consider a path integral representation of 
the model.  The Euclidean action may be written
\begin{eqnarray}
\label{S}
S & = & \epsilon \sum_\tau 
\left( H_w + H_{\rm bond} + H_{\rm ring} \right) \nonumber \\
& - & J^\tau \sum_{r\tau}
\cos(\theta_{r\tau + 1} - \theta_{r\tau} + 2\lambda_{r\tau})  
\nonumber \\
& - & K^\tau \sum_{\la rr'\ra \tau} 
\cos(\phi_{rr', \tau+1} - \phi_{rr', \tau} 
     + \lambda_{r\tau}+ \lambda_{r'\tau}) \nonumber \\
& - & 2 W^\tau \sum_{r\tau}\cos(\lambda_{r\tau}) ~. 
\end{eqnarray}
Here, $\lambda_{r\tau} \in [0, 2\pi)$ is a phase variable living on 
the temporal links.  To arrive at this form of the action, we first 
decoupled the Hubbard-$U$ term in the path integral and replaced 
all the Villain forms by cosines.  The lattice spacing in the time 
direction is $\epsilon$, and the various couplings are
$J^\tau = 1/(2 u_b \epsilon)$, $K^\tau = 1/(2 u_\psi \epsilon)$,
and $W^\tau = 1/(4 U \epsilon)$.

The action represents a {\em classical} three-dimensional $XY$ model 
with a global $U(1)$ symmetry.  As all the Boltzmann weights are 
positive, we may analyse the phase diagram of the model using direct
Monte Carlo simulations.  To avoid unimportant complications, we will 
consider a particular choice of coupling constants where 
$J^\tau = 2 \epsilon J_{\rm bond} \equiv J$, 
$K^\tau = 2 \epsilon K_{\rm ring} \equiv K$, and
$W^\tau = 2 \epsilon w \equiv W$.
Our choices of couplings $J$, $K$, and $W$ are such that the resulting 
classical statistical mechanical system is relatively 
isotropic in space-time.\cite{statmech_system}  

\begin{figure}
\epsfxsize=\columnwidth
\centerline{\epsffile{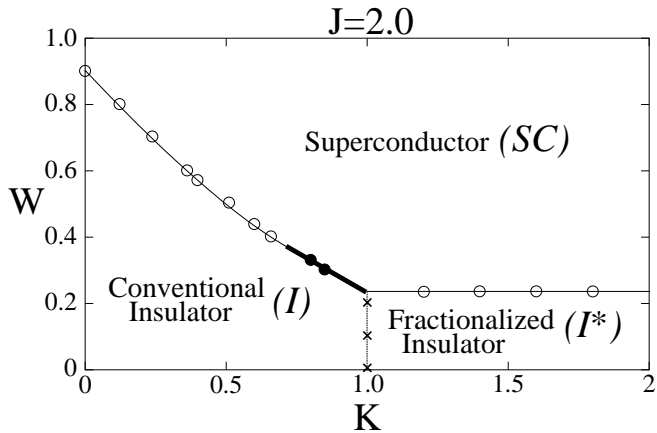}}
\vspace{0.15in}
\caption{Phase diagram of the classical model Eq.~(\ref{S})
at fixed $J=2.0$.  We label the phases using the language
of the original quantum model Eq.~(\ref{H_boson}):
Superconductor $SC$ is an $XY$ ordered phase, while insulators 
$\cal I$ and $\cal I^*$ are two magnetically disordered but 
topologically distinct phases of the classical problem.
}
\label{boson_phased}
\end{figure}

When $W = 0$, the model is easily seen to reduce to the 
{\em classical} $3$D compact QED coupled to a charge~$2$ scalar. 
This has two phases neither of which has $XY$ order 
(which implies insulating behaviour for the original quantum model) 
but which are topologically distinct.   
Turning on a small non-zero $W$ does not induce $XY$ order, 
but preserves the topological distinction between the two phases. 
Upon increasing $W$, there is eventually a transition to an 
$XY$ ordered phase.  Thus, we expect that a cut through the phase 
diagram in the $K$-$W$ plane for large but finite $J$ will look as 
in Fig.~\ref{boson_phased}.

\narrowtext

We verify this expectation by direct simulations of the classical 
model Eq.~(\ref{S}) on cubic lattices of sizes up to $12^3$ with
periodic boundary conditions.
We use heat-bath local updates and run over $5,000$ Monte Carlo 
iterations per each degree of freedom.
We measure the $XY$ order parameter in the original physical angles 
(e.g., $M = \sum_j e^{i\theta_j}$)
and the superfluid stiffness $\rho_s$ associated with the direct 
boson hopping $w$ ($\rho_s$ is defined in a standard way---see, e.g., 
Refs.~\onlinecite{XY3dFSS}).  Both quantities can be used to identify 
transitions into the superfluid phase.  We also measure 
the specific heat of the classical system; this serves
as an unbiased indication of the thermodynamic phase transitions 
and their order.  From these studies performed at fixed 
moderately large $J=2.0$, we obtain the phase diagram shown in
Fig.~\ref{boson_phased}, where we find three phases: 
an $XY$ ordered phase ($SC$) and two distinct disordered phases 
($\cal I$ and $\cal I^*$).  We should point out one detail about our 
scans through the parameter space:  To perform an accurate study 
of the $SC$ to $\cal I^*$ transition, we always start from a fully 
ordered state inside the $SC$ phase.  We found that if we start from 
a completely disordered state in the ${\cal I}^*$ phase, 
the system often traps a vison and subsequently a vortex when going 
into the $SC$ phase (see also our discussion below), 
which significantly affects the measurements in our systems.

We analyze the transitions using finite-size scaling.
The ${\cal I}$ to $SC$ transition for small $K \leq 0.6$
and the ${\cal I^*}$ to $SC$ transition exhibit $3$D~$XY$
critical behavior; these are shown with open symbols in
Fig.~\ref{boson_phased}.  For example, we can use the 
finite-size scaling relation, $\rho_s L = g(L^{1/\nu}t)$,
to locate the transitions and determine the correlation
length exponent $\nu$.  Using this standard procedure,
we also observe an important distinction between the two 
disordered phases:  The universal value 
$(\rho_s L)_{\rm crit}$ at the ${\cal I^*}$ to $SC$ transition
is found to be one-fourth that at the ${\cal I}$ to $SC$
transition, consistent with the charge fractionalization
in the ${\cal I^*}$ phase.

The ${\cal I}$ to $SC$ transition for larger values of $K$
approaching the ${\cal I^*}$ phase, $0.7 \leq K \leq 1.0$,
seems to be first order; this is indicated with filled symbols
and a heavy line in the same figure.  Our evidence for this
is the observed strong sharpening of the specific heat peak 
for the larger systems, with the maximum value growing
very strongly with the system size.

The ${\cal I}$ to ${\cal I^*}$ transition 
(marked by crosses in Fig.~\ref{boson_phased}) is most easily 
identified by observing the specific heat.  This is a true
thermodynamic transition, but is not accompanied by any conventional 
ordering.  The finite-size scaling of the specific heat peak is 
consistent with the $3$D~Ising universality class.

\begin{figure}
\epsfxsize=\columnwidth
\centerline{\epsffile{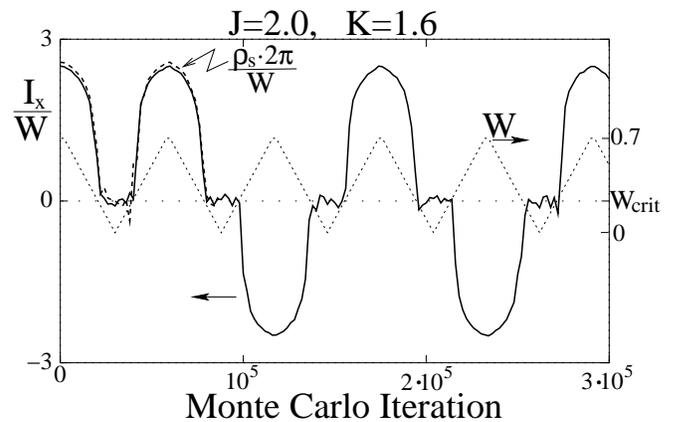}}
\vspace{0.15in}
\caption{Flux trapping experiment at $J=2.0$, $K=1.6$.
The system (of size $8^3$) is prepared with a single vortex in 
the annulus encircled by the $L_x$, deep in the $SC$ phase 
($W=0.7$, cf.~Fig.~\ref{boson_phased}).  The vortex is detected 
by measuring the circulation $I_x$ of the superfluid current 
(solid line).  The system is cycled between the $SC$ and 
$\cal I^*$ phases.  The ``saw-tooth'' dotted line is the MC time 
variation of $W$ drawn so that the critical $W_{\rm crit}$ 
coincides with the zero of $I_x$.  For a trapped vortex, 
the magnitude of $I_x$ is set by the superfluid stiffness
and is expected to be $\approx \rho_s \!\cdot\! 2\pi$.
The latter is shown with a dashed line for the first
two cycles; the fact that the two quantities coincide 
indicates that the vortex remains trapped across
the transition.
}
\label{vtrap}
\end{figure}

To illuminate the topological order in the ${\cal I^*}$ phase,
we perform flux trapping ``experiments'' as described in
Ref.~\onlinecite{toexp} 
(see also Ref.~\onlinecite{SedScaSug}).  We summarize these
experiments in Fig.~\ref{vtrap}.  The system is prepared 
deep in the $SC$ phase with one vortex inside the annulus
encircled by the periodic $L_x$; the physical angles 
$\theta$ and $\phi$ accumulate phase $2\pi$ going 
around the $L_x$, and there is a superfluid current with
circulation 
$I_x \equiv \oint \vec j \!\cdot\! d\vec l 
\approx \rho_s \!\cdot\! 2\pi$ 
in this direction.
As we decrease $W$ towards the ${\cal I^*}$ phase,
the vortex remains trapped all the way to the transition, 
and the magnitude of the superfluid current is set by $\rho_s$.
In the ${\cal I^*}$ phase, the superfluid current is, of 
course, zero, but when we cycle the system back into the 
$SC$ phase, the superfluid current reappears with 
full initial strength but with a random sign.
For comparison, if we create a double vortex and perform
a similar $SC$-${\cal I^*}$ cycle (not shown) on similar
time scales, the double vortex ``tunnels out'' before we 
reach the ${\cal I^*}$ phase and never reappears again.  
Similarly, a single or a double vortex created in the $SC$
phase both disappear when we approach the ${\cal I}$ phase
and never reappear again upon subsequent $SC$-${\cal I}$ 
cycling.

\narrowtext

In terms of the effective degrees of freedom of the ${\cal I^*}$
phase, the physical vortex is formed by a $\pi$ vortex in the 
chargon field and a vison.  Bringing the system into the 
${\cal I^*}$ phase, the vison remains gapped and is trapped
in the annulus.  Cycling the system back into the $SC$ phase, 
the vison binds a $\pi$ vortex in the chargon field, thus creating
a physical vortex in the annulus but with a random sign.

\subsection{Effective field theory}
\label{eft}
We now provide a mapping of the model Hamiltonian to a $Z_2$ 
gauge theory that will make obvious the results mentioned before. 
In addition, this yields an explicit derivation of the effective 
field theory for the fractionalized phase. 

Consider the Hamiltonian Eq.~(\ref{H_boson}).
To bring out the possibility of a fractionalized phase, 
define the operators $b_{cr}= e^{i\theta_{cr}}$ and 
$\tilde\Psi_{rr'} = e^{i\tilde\phi_{rr'}}$ through
\begin{equation}
b_{cr}^\dagger  =  s_r e^{i\theta_r / 2} ~, \quad\quad
\tilde\Psi_{rr'}^\dagger =  \Psi_{rr'}^\dagger b_{cr} b_{cr'} ~.
\label{chargon}
\end{equation}
Here $s_r = \pm 1$ so that $\theta_{cr} \in [0, 2\pi)$. 
The field $b_{cr}$ may be thought of as the ``square root'' of 
the operator $b_r$ and carries charge $\frac{q_b}{2}$ and may be 
interpreted as a chargon operator. 
The field $\tilde\Psi_{rr'}$ on the other hand is charge neutral.   

Clearly, the boson number $N_r$ is conjugate to $\theta_{cr}$ and 
commutes with $\tilde\phi_{rr'}$, while $n_{rr'}^\psi$ is conjugate 
to $\tilde\phi_{rr'}$ and commutes with $\theta_{cr}$:
\begin{eqnarray}
[\theta_{cr}, N_r]  & = & i ~, \quad\quad 
[\tilde\phi_{rr'}, N_r] = 0 ~, \nonumber \\ \,
[\tilde\phi_{rr'}, n_{rr'}^\psi]  & = & i ~, \quad \quad 
[\theta_{cr}, n_{rr'}^\psi] = 0 ~.
\label{commutators}
\end{eqnarray}
We can now write the Hamiltonian in terms of 
$(\theta_{cr}, N_r, \tilde\phi_{rr'}, n_{rr'}^\psi)$,
rather than the original variables.
However, to recover the original physical Hilbert space,
we need to impose the constraint 
\begin{equation}
e^{i\pi(N_r - \sum_{r'\in r} n_{rr'}^{\psi})} = 1 ~;
\label{constraint}
\end{equation}
this ensures that $n_r^b = (N_r-\sum_{r'\in r} n_{rr'}^{\psi})/2$ 
is an integer (original $b$-boson) number operator. 

Making this (exact) change of variables, we obtain for the parts of
the Hamiltonian Eq.~(\ref{H_boson})
\begin{eqnarray}
H_w & = & - w \sum_{r, r'\in r} 
  (b_{cr}^\dagger \tilde\Psi_{rr'} b_{cr'}+ h.c.) ~,\\ 
H_{\rm bond} & = & - J_{\rm bond} \sum_{\la rr' \ra} 
  \left[ (\tilde\Psi_{rr'}^\dagger)^2 + h.c. \right] ~, \\
H_{\rm ring} & = & - K_{\rm ring} \sum_\Box 
  ( \tilde\Psi_{12}^\dagger \tilde\Psi_{23} 
    \tilde\Psi_{34}^\dagger \tilde\Psi_{41} + h.c. ) ~.
\end{eqnarray}

Note that the $H_{\rm bond}$ term acts as an Ising anisotropy on the 
$\tilde\Psi_{rr'}$ field.  Considerable simplification is possible in the 
limit of large $J_{\rm bond}$ and small $u_b$ to which we now specialize. 
The potential ``seen'' by the phase $\tilde\phi_{rr'}$ has two deep 
equivalent minima $\tilde\phi_{rr'}=0$ or $\pi$, which we label by
$\sigma_{rr'}^z = e^{i\tilde\phi_{rr'}}=\pm 1$.
The kinetic term $(n_{rr'}^\psi)^2$ causes tunneling between 
the two wells.  At each link, there are two low-energy states separated 
from all other states by a gap, leading to an effective $2$-state system.  
In the $\sigma_{rr'}^z$ basis, we identify
$e^{i\pi n_{rr'}^\psi} = \sigma_{rr'}^x$ since this operator
translates $\tilde\phi_{rr'}$ by $\pi$.
Also, the kinetic term $u_\psi (n_{rr'}^\psi)^2$ is replaced by
an effective transverse field $h \sigma_{rr'}^x$.
In this large $J_{\rm bond}$ limit, the effective Hamiltonian becomes
\begin{eqnarray}
\label{H_chgon}
H_{\rm ch}[b_c,\sigma] & = & - w \sum_{\la rr'\ra} 
(\sigma_{rr'}^z b_{cr}^\dagger b_{cr'} + h.c.) 
+ U\sum_r N_r^2 \nonumber \nonumber \\
&  & - 2 K_{\rm ring}\sum_\Box 
\sigma_{12}^z \sigma_{23}^z \sigma_{34}^z \sigma_{41}^z
-h\sum_{\la rr' \ra} \sigma_{rr'}^x ~,
\end{eqnarray}
while the constraint Eq.~(\ref{constraint}) is written as
\begin{equation}
(-1)^{N_r} \prod_{r'\in r} \sigma_{rr'}^x = 1
\end{equation}
at each site $r$.

This effective model is precisely the quantum problem of chargons 
coupled to a fluctuating $Z_2$ gauge field in two dimensions  
introduced and analyzed in Ref.~\onlinecite{z2long}.  
This model is known to have a phase diagram of the kind shown in 
Fig.~\ref{boson_phased}.  In particular, there is an insulating 
fractionalized phase where the chargon fields are deconfined and 
there is a gapped vison (which occurs for large 
$K_{\rm ring}$, and small $w$).

\subsection{Generalization to arbitrary commensurate filling}
Our results are readily generalized to arbitrary commensurate values of 
the total average number of bosons per unit cell. 
Consider a modification of the Hamiltonian where the $U$ term is 
replaced by 
\begin{equation}
U(N_r - N_0)^2 
\end{equation}
with $N_0$ a constant.  At such commensurate density 
(rational values of $N_0$), insulating phases of the bosons will be 
possible.  Again, in the limit of large $J_{\rm bond}, K_{\rm ring}, U$ 
and small $u_b, u_\psi, w$, this insulator will be fractionalized. 
Conventional (i.e.~non-fractionalized) insulating states are of course 
possible in other limits.  All of these cases are readily studied using 
the methods of the previous subsection.  Indeed, a non-zero $N_0$ is 
trivially incorporated with no essential change leading to an effective 
Hamiltonian Eq.~(\ref{H_chgon}) but with the modified $U$ term above.  
As a special case of some interest, consider $N_0 = 1$. 
The resulting model has previously been suggested\cite{gof,ParSach_long} 
as an effective model of frustrated easy-plane spin-$1/2$ quantum 
antiferromagnets in two dimensions.  
A recent study by Park and Sachdev\cite{ParSach_short}
explicitly demonstrates the presence of the expected two insulating 
phases: a bond density wave crystal with confined excitations
and a fractionalized phase.
The fractionalized phase is more stable in this $N_0=1$ case
due to additional frustration coming from the Berry phase terms.

\section{Models for spin-charge separation}
In this section, we generalize the models of Section~\ref{Sec:boson} 
to construct models that display spin-charge separated phases. 
We follow the route to spin-charge separation explored 
in Ref.~[\onlinecite{NLII,z2long}] by considering models of electrons 
coupled to superconducting phase fluctuations.  
These may be thought of as models of spin-$1/2$ charge $e$ electrons 
interacting with spin-$0$ charge $2e$ Cooper pairs.  As shown below, 
the independent excitations of the spin-charge separated phase are 
(i) a spin-$0$ charge $e$ chargon, 
(ii) a spin-$1/2$ charge $0$ spinon, and 
(iii) a spinless charge neutral $Z_2$ vortex---the vison. 
When either a chargon or spinon is taken all the way around a vison, 
the system acquires a phase of $\pi$.  This structure is exactly what 
is expected on the basis of the effective field theories of stable 
spin-charge separated phases.  Indeed, as shown below, it is possible 
to provide an explicit derivation of the effective field theory as the 
correct description of our models in appropriate limits. 

In the models presented below, the spinons are fermions while the 
chargons are bosons.  An important property of the spinons is that 
their number is not conserved.  There are ``pairing'' terms in the 
Hamiltonian describing the spinon dynamics.  Different spin-charge 
separated phases obtain based on the pairing symmetry of the spinons. 
Below we will discuss two different pairing symmetries as illustrative 
examples.

\subsection{Model for $d$-wave paired spinons}
Consider the following model:
\begin{eqnarray}
\label{H_dwave}
H_{d{\rm wave}} & = & H_t + H_{\Delta} + H_w + H_{\rm bond} 
+ H_{\rm ring} + H_u ~, \\
H_t & = & - t\sum_{\la rr' \ra} \left(c^{\dagger}_{r\alpha} c_{r'\alpha} +
h.c. \right) ~, \\
H_{\Delta} & = & \sum_{\la rr' \ra} \Delta_{rr'}
\left[\Psi^\dagger_{rr'} 
      \left( c_{r\ua} c_{r'\da} - c_{r\da} c_{r'\ua} \right) 
      + h.c. \right] ~, \\
H_u & = & u_{\psi} \sum_{\la rr' \ra} (n^\psi_{rr'})^2 
+ U \sum_r (N_r - N_0)^2 ~.
\end{eqnarray}
Here, $c_{r\alpha}$ represents the destruction operator for an electron 
at site $r$ and spin $\alpha$.  The electron is taken to have charge $e$. 
The operator $\Psi_{rr'}$ may, in this model, be considered a Cooper pair 
living on the bonds of the lattice.  In addition, these Cooper pairs on 
the bonds are coupled to other Cooper pair degrees of freedom $b_r$
residing on the sites of the lattice.  The corresponding boson-only terms 
$H_w$, $H_{\rm bond}$, and $H_{\rm ring}$ are the same as before. 
The operator $N_r$ is defined through
\begin{equation}
N_r = 2 n_r^b + \sum_{r' \in r} n_{rr'}^\psi 
+ \sum_\alpha c^\dagger_{r\alpha} c_{r\alpha} ~.
\end{equation}
Clearly, the total charge $Q_{\rm tot} = e \sum_r N_r$. 
The number $N_0$ is a constant that sets the average charge per site. 
We take the ``pairing amplitude'' $\Delta_{rr'}$ to have $d_{x^2 - y^2}$ 
symmetry. 

If $U$ is large, the system will be in an insulating phase
(for commensurate density). 
The properties of this insulator depend on the values of the other 
parameters in the Hamiltonian.  In particular, for large 
$J_{\rm bond}, K_{\rm ring}$, 
we argue that the insulator will be spin-charge separated. 
The spinons are fermionic and have $d_{x^2- y^2}$ pairing symmetry. 

We proceed as before and define the chargon field $b_{cr}$ 
and the neutral field $\tilde \Psi_{rr'}$ through Eqs.~(\ref{chargon}).
It will also be extremely convenient to define a spinon field 
$f_{r\alpha}$ through
\begin{equation}
c_{r\alpha} = b_{cr} f_{r\alpha} ~.
\end{equation}
As before, the total charge associated with each site $N_r$ is 
conjugate to the chargon phase $\theta_{cr}$ and commutes with both 
$\tilde\Psi_{rr'}$ and $f_{r\alpha}$:
\begin{equation}
[\theta_{cr}, N_r] = i ~, \quad\quad
[\tilde\Psi_{rr'}, N_r] = [f_{r\alpha}, N_r] = 0 ~.
\end{equation}
As expected, the $f_{r\alpha}$ fields are formally charge neutral.
Equations~(\ref{commutators}) also continue to hold.
We further have
\begin{equation}
[f_{r\alpha}, n_{rr'}^\psi] = 0
\end{equation}
and the equality 
$c^\dagger_{r\alpha} c_{r\alpha} = f^\dagger_{r\alpha} f_{r\alpha}$
We may work with the set of variables 
$(b_{cr}, N_r, \tilde\Psi_{rr'}, n_{rr'}^\psi, f_{r\alpha})$
instead of the original set 
$(b_r, n_r^b, \Psi_{rr'}, n_{rr'}^\psi, c_{r\alpha})$. 
As with the boson-only models, this requires imposing a constraint on 
the Hilbert space which now takes the form
\begin{equation}
(-1)^{N_r - \sum_{r' \in r} n_{rr'}^\psi - f_r^\dagger f_r} = 1 ~.
\end{equation}
Continuing with the same steps as in Section~\ref{Sec:boson}, 
we find that in the large $J_{\rm bond}$ limit, the Hamiltonian 
reduces to the following:
\begin{eqnarray}
H & = & 
H_t + H_\Delta + H_{\rm ch}[b_c, \sigma] ~, \\
H_t & = & -t \sum_{\la rr' \ra}
\left(f_r^\dagger f_{r'} b_{cr}^\dagger b_{cr'} + h.c. \right) ~, \\
H_\Delta & = & \sum_{\la rr' \ra} \Delta_{rr'} 
\left[ \sigma_{rr'}^z (f_{r\ua} f_{r'\da} - f_{r\da} f_{r'\ua}) + h.c.
\right] ~.
\end{eqnarray}
$H_{\rm ch} [b_c, \sigma]$ is the same as before, Eq.~(\ref{H_chgon}). 
The constraint reduces to
\begin{equation}
(-1)^{N_r - f_r^\dagger f_r} \prod_{r' \in r} \sigma_{rr'}^x = 1~.
\end{equation}

We now argue that for $N_0$ an integer, there is a stable spin-charge 
separated phase.  We first note that the Hamiltonian above describes a 
$Z_2$ gauge theory of spinons and chargons coupled to the $Z_2$ gauge 
field.  As such, for large $K_{\rm ring} \gg h$, it's structure is almost 
identical to the effective theory of a spin-charge separated phase 
of Ref.~\onlinecite{z2long}.  The main difference is in the nature 
of the spinon hopping term (the term $H_t$) which seems to couple 
together the spinons and the chargons.  However, this is readily seen 
to be an unimportant difference.

First, consider the limit of small $t, w$ (but $t \ll w$) 
at large repulsion $U$.  In this limit, the chargons will lock into a 
Mott insulating phase (at integer $N_0$). 
At $t = w = 0$, the chargon number will be fixed at $N_0$ per site. 
Going slightly away from this limit, we may treat both $H_t$ and the 
chargon hopping term in perturbation theory to eliminate virtual 
charge fluctuations.  The result will be an effective Hamiltonian 
describing the spinon and gauge degrees of freedom. 
To second order, the generated terms take the form
\begin{equation}
-\sum_{\la rr' \ra} 
\frac{  \hat V^\dagger \hat V + \hat V \hat V^\dagger } {2 U} ~,
\end{equation}
with $\hat V = t f_r^\dagger f_{r'} + w \sigma_{rr'}^z$.
Expanding, we get two non-trivial terms: the first is simply spinon 
hopping coupled to the $Z_2$ gauge field,  while the second is a spinon 
four fermion interaction.  The effective Hamiltonian then becomes
\begin{eqnarray}
H & = & H_{{\rm sp}, t} + H_{\rm sp, int} + H_{\Delta} 
+ H_{\rm IGT}[\sigma] ~,\\
H_{{\rm sp}, t} & = & 
-t_{\rm sp} \sum_{\la rr' \ra} \sigma_{rr'}^z 
\left( f_r^\dagger f_{r'} + h.c. \right) ~,\\
H_{\rm sp, int} & = & -\lambda \sum_{\la rr' \ra} 
\left[ (f_r^\dagger f_{r'}) (f_{r'}^\dagger f_r) + 
       (f_{r'}^\dagger f_r) (f_r^\dagger f_{r'}) \right] ~.
\end{eqnarray}
Here the spinon hopping $t_{\rm sp} = tw/U$ and the spinon 
interaction strength $\lambda = t^2/(2U)$.
Furthermore, in this large $U$ limit, the constraint simply reduces to 
\begin{equation}
(-1)^{f_r^\dagger f_r + N_0} = \prod_{r' \in r} \sigma_{rr'}^x ~.
\end{equation}
As a function of $K_{\rm ring}$, this Hamiltonian undergoes a 
deconfinement transition.  
In particular, for large $K_{\rm ring}$, the fluctuations of the 
gauge field may be ignored and the spinons are free to propagate. 
The nature of the spinon dispersion is easily found by considering 
the limit $K_{\rm ring} = \infty$.  In this limit, we may set 
$\sigma_{rr'}^z = +1$ on every bond.  The quadratic part of the 
spinon Hamiltonian is then formally the same as that describing 
non-interacting quasiparticles in a $d_{x^2 - y^2}$ superconductor, 
and therefore describes gapless nodal spinons.  The spinon interaction 
is a formally irrelevant perturbation at this free spinon theory.  
As we are specifically in the limit that $t \ll w$, 
we have $\lambda \ll t_{\rm sp}$---thus the interaction term may be 
safely ignored.  Making $K_{\rm ring}$ finite also only leads to 
irrelevant perturbations to the free spinon theory so that the long 
distance spin physics of the spin-charge separated phase is described by 
nodal fermionic spinons. 

The argument above considered the limit of large $U, K_{\rm ring}$ 
but small $t, w$.  
It is also instructive to consider the limit $t,w \gg U$. 
In this limit, the chargons are expected to Bose condense leading to 
an ordinary $d_{x^2 - y^2}$ superconductor. 
The long distance physics of this superconducting phase is readily 
captured by a continuum theory which keeps a continuum chargon 
phase field and the nodal spinons.  The electron kinetic energy term 
$H_t$ is then readily written as a spinon kinetic energy modified by 
the usual ``Doppler shift'' term coupling the gradient of the phase 
(the superflow) to a bilinear in the spinons.  
Vortices are permitted in this phase and have flux quantized in 
multiples of $hc/2e$.  In the large $K_{\rm ring}$ limit, it is easy to 
see that the core energy of an $hc/2e$ vortex will include a 
contribution proportional to $K_{\rm ring}$.  On the other hand, the core 
energy of $hc/e$ vortices does not diverge as $K_{\rm ring}$ goes to 
infinity. 
Now consider decreasing $w$ to induce a transition to the insulator. 
At large $K_{\rm ring}$, it is clear that this will occur due to 
proliferation of 
$hc/e$ vortices rather than due to $hc/2e$ vortices.  Following the 
general arguments in Ref.~[\onlinecite{NLII,z2long}], we will get a 
spin-charge separated phase.  Note that, as argued in 
Ref.~\onlinecite{NLII}, the Doppler shift term coupling the 
chargons and spinons is formally irrelevant, and one obtains a 
nodal liquid phase.

\subsection{Model for $s$-wave paired spinons}
It is straightforward to modify the model above to obtain one that 
stabilizes a spin-charge separated phase with $s$-wave paired 
fermionic spinons with a spin gap.  We merely modify the pairing 
term above to 
\begin{equation}
H_\Delta = \Delta\sum_r b_r^\dagger c_{r\ua} c_{r\da} + h.c.
\end{equation}
Proceeding exactly as above, it is easily established that for 
large $J_{\rm bond}, U, K_{\rm ring}$, such a spin-charge separated 
phase is indeed realized.

\subsection{Model for spin-charge separated magnetically ordered 
phases}
The effective field theories for spin-charge separated phases 
strongly suggest the theoretical possibility of spin-charge separation 
coexisting with magnetic long range order in a quantum phase. 
In this subsection, we show how the models above may be readily 
generalized to stabilize such phases.  Consider a system consisting of 
two layers and a Hamiltonian of the form
\begin{equation}
H = H^{(1)} + H^{(2)} + H^{(12)} ~.
\end{equation}
Here $H^{(1)}, H^{(2)}$ refer to parts of the Hamiltonian that 
depend only on the degrees of freedom residing in layers $1$ and $2$
respectively.  The interactions between the two layers are contained 
in the term $H^{(12)}$.  We assume that layer $1$ consists of a 
square lattice of Heisenberg spins with $S = 1/2$ described by the 
Heisenberg antiferromagnetic model
\begin{equation}
H^{(1)} = J_1 \sum_{\la rr' \ra} \mb S_{1r} \cdot \mb S_{1r'} ~.
\end{equation}
We assume that layer $2$ is described by the Hamiltonian 
$H_{d\rm wave}$ in Eq.~(\ref{H_dwave}) above, and that the 
interaction between the two layers is given by
\begin{equation}
H^{(12)} = J_\perp \sum_r 
\mb S_{1r} \cdot (c_{2r}^\dagger \bfgr\sigma c_{2r} )
\end{equation}
with $J_\perp \ll J_1$.  We assume that at $J_\perp = 0$, 
the layer $2$ is in it's spin-charge separated (and hence 
topologically ordered) phase.  In this limit, layer $1$ will order 
antiferromagnetically.  Turning on a weak coupling $J_\perp$ 
will induce antiferromagnetic ordering in layer $2$, 
but cannot destroy the vison gap.  Consequently, the full 
model Hamiltonian will be in a phase that has magnetic long range 
order but nevertheless is spin-charge separated.

\section{Summary}
In this paper, we have discussed several concrete examples of 
microscopic models in two spatial dimensions that display quantum phases 
with fractionalized excitations.  These models possess no special 
symmetries other than those associated with global charge or spin 
conservation and also have only short ranged interactions, 
and thus confirm that fractionalization is a theoretically acceptable 
possibility for strongly interacting many particle systems in spatial 
dimension bigger than one.  These models explicitly realize earlier 
field theoretic descriptions of fractionalization phenomena. 

A number of generalizations of our results are possible. 
Our models are easily generalized to arbitrary spatial dimension, 
and provide concrete examples of fractionalized phases in any spatial 
dimension $d > 1$.  For spin-charge separated phases of electronic 
systems, we have chosen to describe models with fermionic spinons and 
bosonic chargons.  Following the ideas in Ref.~\onlinecite{gof}, 
these are readily modified to construct spin-charge separated phases with 
fermionic chargons and bosonic spinons (at least with easy plane spin 
anisotropy).  An additional upshot of our results is the construction of 
topologically ordered {\em classical} $3$D $XY$ 
models\cite{statmech_system}. 

Finally, we mention that quantum phases with topological order have 
also been suggested\cite{Kit} to be suitable states of interest to 
quantum computation. 
The topological structure naturally protects the system from decoherence. 
This very preliminary application\cite{NaySht,Iof} may also benefit from 
the results in this paper. 

We thank Leon Balents, Matthew Fisher, and Subir Sachdev for useful 
conversations. 
This work was supported by the MRSEC program of 
the National Science Foundation under grant DMR-9808941.


\begin{references}
\bibitem{statmech_system}  Note that the $W^\tau$ term in 
the action~(\ref{S}) suppresses the fields $\lambda_{r\tau}$ 
and thus promotes the correlations of the physical angles 
$\theta_{r\tau}$ and $\phi_{rr' \tau}$ in the time direction.  
In this sense, it is similar to the spatial boson hopping term $w$.
Here, we also mention a particularly simple classical model that 
is qualitatively very similar (but not identical) to the 
one described by the action Eq.~(\ref{S}).  This $3$D isotropic model 
has angles $\theta_i$ and $\phi_{ij}$ living on the corner and 
bond-centered sites of a cubic lattice, with the classical energy
\begin{eqnarray*}
& - & W \sum_{i, j\in i} \cos(\theta_i - \phi_{ij}) 
- J \sum_{\la i j \ra} \cos(\theta_i + \theta_j - 2\phi_{ij}) \\
& - & K \sum_\Box \cos(\phi_{12} - \phi_{23} + \phi_{34} - \phi_{41}) 
~.
\end{eqnarray*}
It is remarkable that this simple $XY$-like model has two 
magnetically disordered phases, one of which is topologically ordered!
In the main text, we have chosen to simulate a somewhat more
complicated looking action Eq.~(\ref{S}) so as not to divert
the reader's attention from the quantum Hamiltonian 
Eq.~(\ref{H_boson}).
\end{references}
\end{document}